\documentclass[11pt,prl,reprint,superscriptaddress,nofootinbib]{revtex4-1}

\usepackage{amsmath,amssymb,amsfonts, braket}
\usepackage{graphicx}
\usepackage{bm}
\usepackage[hidelinks]{hyperref}
\usepackage[usenames]{color}
\usepackage[T1]{fontenc}
\usepackage{booktabs}
\usepackage{siunitx}

\makeatletter 
\renewcommand\onecolumngrid{
\do@columngrid{one}{\@ne}%
\def\set@footnotewidth{\onecolumngrid}
\def\footnoterule{\kern-6pt\hrule width 1.5in\kern6pt}%
}
\renewcommand\twocolumngrid{
        \def\footnoterule{
        \dimen@\skip\footins\divide\dimen@\thr@@
        \kern-\dimen@\hrule width.5in\kern\dimen@}
        \do@columngrid{mlt}{\tw@}
}%
\makeatother 

\newcommand{\er}{Er$^{3+}$ }
\newcommand{\ermgo}{Er$^{3+}$:MgO}

\begin{document}

\title{Strong Purcell enhancement of an optical magnetic dipole transition}

\author{Sebastian P.\ Horvath}
\thanks{These authors contributed equally to this work.}
\affiliation{Department of Electrical and Computer Engineering, Princeton University, Princeton, NJ 08544, USA\looseness=-1}
\author{Christopher M.\ Phenicie}
\thanks{These authors contributed equally to this work.}
\affiliation{Department of Electrical and Computer Engineering, Princeton University, Princeton, NJ 08544, USA\looseness=-1}
\author{Salim Ourari}
\thanks{These authors contributed equally to this work.}
\affiliation{Department of Electrical and Computer Engineering, Princeton University, Princeton, NJ 08544, USA\looseness=-1}
\author{Mehmet T.\ Uysal}
\affiliation{Department of Electrical and Computer Engineering, Princeton University, Princeton, NJ 08544, USA\looseness=-1}
\author{\\Songtao Chen}
\thanks{Present address: Department of Electrical and Computer Engineering, Rice University, Houston, Texas 77005, USA}
\affiliation{Department of Electrical and Computer Engineering, Princeton University, Princeton, NJ 08544, USA\looseness=-1}
\author{\L{}ukasz Dusanowski}
\affiliation{Department of Electrical and Computer Engineering, Princeton University, Princeton, NJ 08544, USA\looseness=-1}
\author{Mouktik Raha}
\thanks{Present address: AWS Center for Quantum Networking, Boston, Massachusetts 02135, USA}
\affiliation{Department of Electrical and Computer Engineering, Princeton University, Princeton, NJ 08544, USA\looseness=-1}
\author{Paul Stevenson}
\thanks{Present address: Department of Physics, Northeastern University, Boston, Massachusetts 02115, USA}
\affiliation{Department of Electrical and Computer Engineering, Princeton University, Princeton, NJ 08544, USA\looseness=-1}
\author{Adam T.\ Turflinger}
\affiliation{Department of Electrical and Computer Engineering, Princeton University, Princeton, NJ 08544, USA\looseness=-1}
\author{\\Robert J.\ Cava}
\affiliation{Department of Chemistry, Princeton University, Princeton, NJ 08544, USA\looseness=-1}
\author{Nathalie P.\ de Leon}
\affiliation{Department of Electrical and Computer Engineering, Princeton University, Princeton, NJ 08544, USA\looseness=-1}
\author{Jeff D.\ Thompson}
\email{jdthompson@princeton.edu}
\affiliation{Department of Electrical and Computer Engineering, Princeton University, Princeton, NJ 08544, USA\looseness=-1}

\begin{abstract}
Engineering the local density of states with nanophotonic structures is a powerful tool to control light-matter interactions via the Purcell effect. At optical frequencies, control over the electric field density of states is typically used to couple to and manipulate electric dipole transitions. However, it is also possible to engineer the magnetic density of states to control magnetic dipole transitions. In this work, we experimentally demonstrate the optical magnetic Purcell effect using a single rare earth ion coupled to a nanophotonic cavity. We engineer a new single photon emitter, Er$^{3+}$ in MgO, where the electric dipole decay rate is strongly suppressed by the cubic site symmetry, giving rise to a nearly pure magnetic dipole optical transition. This allows the unambiguous determination of a magnetic Purcell factor $P_m=1040 \pm 30$. We further extend this technique to realize a magnetic dipole spin-photon interface, performing optical spin initialization and readout of a single  Er$^{3+}$ electron spin. This work demonstrates the fundamental equivalence of electric and magnetic density of states engineering, and provides a new tool for controlling light-matter interactions for a broader class of emitters.
\end{abstract}

\maketitle
\onecolumngrid

The interaction of electromagnetic radiation with matter is of fundamental importance and underlies numerous current and future technologies. In particular, the absorption and emission of light by atomic or molecular transitions has enabled technologies such as the laser, MRI and atomic clocks~\cite{svanberg2012}. The ability to control absorption and emission through engineering the environment is particularly relevant for quantum technologies requiring efficient light-matter interfaces, and has been demonstrated using optical cavities and numerous emitters including atoms and ions \cite{mckeever2004,stute2012,tiecke2014}, quantum dots \cite{arcari2014,hummel2019} and atom-like defects in the solid state \cite{faraon2011,zhong2015,dibos2018,merkel2020}.

Light-matter interaction can take place through multiple processes, including electric dipole (ED), magnetic dipole (MD), or higher order multipole transitions \cite{condon1951}. The natural scale of ED transitions is the largest, and therefore ED transitions are most often targeted for controlling light-matter interactions. However, in certain atoms ED transitions are suppressed, such that higher-order processes become dominant. Many experiments have demonstrated magnetic local density of states (LDOS) engineering in the microwave frequency domain, using metallic or superconducting cavities coupled to spin ensembles \cite{breeze2015} and, recently, individual spins \cite{wang2023}. 

Demonstrating magnetic LDOS engineering in the optical domain is more challenging, as many emitters with significant MD decay pathways have competing decay processes that must be distentangled, such as forced electric dipoles and nonradiative decay. In emitters with mixed MD and ED decay pathways, the relative contributions can be distinguished through the angular spectrum of the emitted radiation~\cite{freed1941}. Furthermore, placing emitters near dielectric interfaces and in thin film structures has enabled small modifications of ED and MD decay rates through the Purcell effect \cite{drexhage1974,lukosz1977,kunz1980,barnes1998,taminiau2012}. However, demonstrating strong modification of the emission of magnetic dipole emitters in the optical domain via magnetic density of states engineering is a long-standing goal \cite{baranov2017,freed1941,drexhage1974,kunz1980,taminiau2012,rolly2012,albella2013,karaveli2013,shafiei2013,decker2013,aigouy2014,hussain2015}.

In this work, we demonstrate strong Purcell enhancement of an optical MD transition in a single Er$^{3+}$ ion using a nanophotonic cavity. This is enabled by engineering a new single photon emitter, Er$^{3+}$:MgO, which by symmetry has a nearly pure MD optical transition at a wavelength of 1540.48 nm. The MD nature of the transition is experimentally confirmed via lifetime measurements, and comparison of the measured fluorescence and absorption spectra with a crystal field model. By evanescently coupling individual Er$^{3+}$ ions to a silicon nanophotonic cavity with a small magnetic mode volume of $V_m = 0.068$~$\mu$m$^3$, we demonstrate a large Purcell enhancement factor of $P_m = 1040 \pm 30$ which can be unambiguously attributed to the magnetic dipole. Additionally, we use the cavity-enhanced MD transition to realize a spin-photon interface. With this, we determine the ground state spin structure and measure the lifetime and coherence time of a single \er spin. This work opens the door to using nanophotonic structures to control MD emission and enables the use of a wider class of atoms and atom-like systems for quantum technologies. 

\begin{figure*}[ht]
    \includegraphics[width=6.5 in]{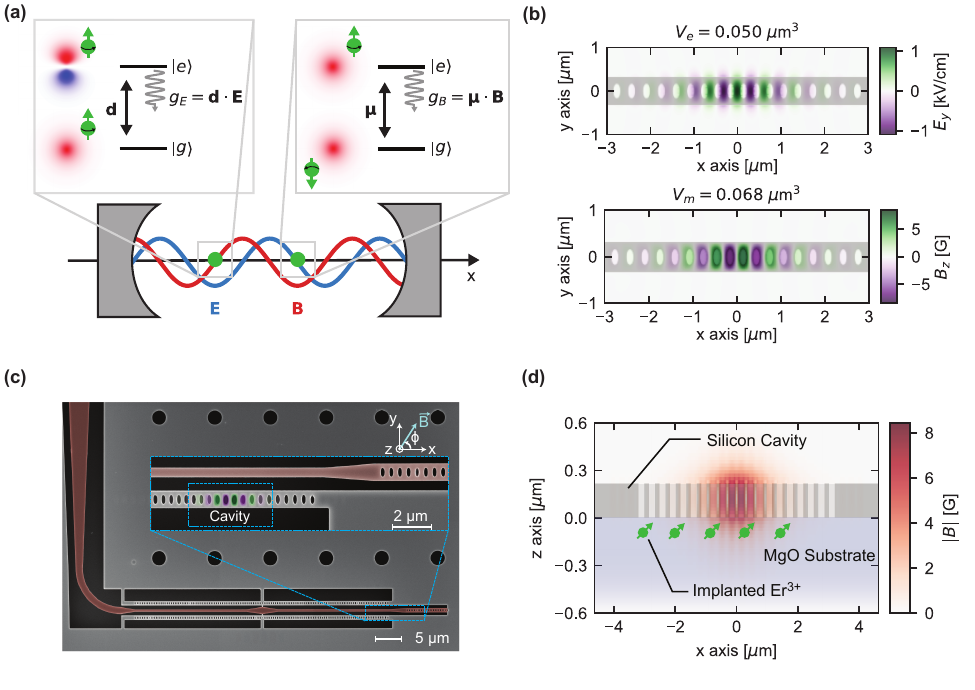}
    \caption{\textbf{Comparison of electric and magnetic dipole coupling to an optical cavity} (a) An emitter with an electric dipole moment $\boldsymbol d$ couples to the electric field of an optical cavity, $\boldsymbol E$, while an emitter with a magnetic dipole moment $\boldsymbol \mu$ couples to the magnetic field, $\boldsymbol B$. In a Fabry-Perot cavity, the $\boldsymbol E$ and $\boldsymbol B$ fields are standing waves with the same transverse profile, but shifted by a quarter wavelength along the cavity axis. (b) In a nanophotonic cavity such as a photonic crystal cavity, the behavior of the $\boldsymbol E$ and $\boldsymbol B$ fields is similar to the Fabry-Perot case, but the mode volumes differ slightly because of the dielectric boundary conditions. For the device shown in panel c), the maximum single-photon magnetic field is 8.4 Gauss. (c) Scanning electron micrograph of the silicon nanophotonic circuit used in this work. A photonic crystal defect cavity (inset) is evanescently coupled to a bus waveguide (red), which is coupled to an optical fiber with a grating coupler (not shown).  (d) Schematic of the silicon nanophotonic crystal cavity bonded to an MgO crystal, which is doped with \er ions at a depth of 0-100 nm using ion implantation. The ions couple evanescently to the magnetic field of the cavity.}
    \label{fig:fig1}
\end{figure*}

Realizing a large magnetic Purcell effect requires two components: a cavity with a large magnetic LDOS, and an emitter with a dominant MD decay pathway. The magnetic LDOS of a cavity can be quantified using the magnetic Purcell factor $P_m$, which is defined analogously to the electric Purcell factor as \cite{breeze2015}:
\begin{equation}
P_{e(m)} = \frac{3}{4\pi^2} \left(\frac{\lambda}{n}\right)^3 \frac{Q}{V_{e(m)}(\boldsymbol r)},
\end{equation}
Here, $Q$ denotes the quality factor of the cavity, while the electric (magnetic) mode volume $V_{e(m)}(\boldsymbol r)$ describes the field strength at the position $\boldsymbol r$ of the emitter. For electric fields, the mode volume is defined by the electric field $\boldsymbol E(\boldsymbol r)$ and relative permittivity $\epsilon_r(\boldsymbol r)$ as:
\begin{equation}
V_e(\boldsymbol r) = \frac{\int \epsilon_r(\boldsymbol r) |\boldsymbol E(\boldsymbol r)|^2 d^3 \boldsymbol r}{\max_r \epsilon_r(\boldsymbol r) |\boldsymbol E(\boldsymbol r)|^2}. 
\end{equation}
For a magnetic dipole transition, the analogous expression is:
\begin{equation}
V_m(\boldsymbol r) = \frac{\int |\boldsymbol B(\boldsymbol r)|^2/\mu_r(\boldsymbol r) d^3\boldsymbol r}{\max_r |\boldsymbol B(\boldsymbol r)|^2/\mu_r(\boldsymbol r)}, 
\end{equation}
where $\boldsymbol B(\boldsymbol r)$ is the the magnetic field of the cavity mode and $\mu_r(\boldsymbol r)$ the relative magnetic permeability. In a simple cavity such as a Fabry-Perot resonator, $V_e$ and $V_m$ are identical for optimally positioned emitters, as the distribution of the $\boldsymbol{E}$ and $\boldsymbol{B}$ fields are the same except for a phase shift along the cavity axis by a quarter wavelength (Fig.~\ref{fig:fig1}(a)).

In this work, we use a dielectric photonic crystal cavity to achieve very small mode volumes. The behavior of the electric and magnetic fields is qualitatively similar to the Fabry-Perot case (Fig.~\ref{fig:fig1}(b)). However, differences in the dielectric boundary conditions for $\boldsymbol{E}$ and $\boldsymbol{B}$ fields results in a slight difference in mode volumes, and from numerical simulations we find $V_e = 0.05$~$\mu$m$^3$and $V_m = 0.068$~$\mu$m$^3$. Therefore, this type of cavity is well-suited for attaining large electric and magnetic Purcell factors, depending on the type of emitter.

To address the second requirement of an emitter with a dominant MD decay pathway, we focus on \er ions. The 1.5\,$\mu$m optical transition in \er connects the $^4I_{15/2}$ and $^4I_{13/2}$ levels; since these have the same parity, an electric dipole transition between them is forbidden in a spherically symmetric environment. In host crystals with low site symmetry, admixtures of $5d$ orbitals can lead to a so-called forced ED transition, which is often the dominant decay pathway \cite{dieke1968}. However, this is forbidden in centrosymmetric environments ~\cite{gorller1998}, which leads us to consider MgO as an \er host.

We incorporate Er into MgO using ion implantation followed by annealing (see Supplementary Information for additional information). To spectroscopically identify the cubic site in MgO, we perform site-selective excitation spectroscopy~\cite{stevenson2022} in a heavily implanted sample ($1\times 10^{14}$ Er/cm$^{2}$, sample A) and measure the lowest few ground ($Z_i$) and excited state ($Y_i$) crystal field levels for six distinct sites. Previous studies of \ermgo{} have found a large number of spectral lines, suggesting that \er incorporates into the crystal in several different configurations \cite{Descamps1964,Ayant1962,Belorizky1966,Borg1970,stevenson2022}. Many do not have cubic point group symmetry, most likely because the \er sits next to a vacancy or interstitial. However, one of the observed sites, with a $Z_1 \rightarrow Y_1$ transition at 1540.48 nm, can be reproduced with a cubic crystal-field model (Fig.~\ref{fig:establishMD}(b)) with only four free parameters, with an r.m.s.\ deviation of 1.6~cm$^{-1}$ (see Supplementary Information for further details).

To provide further confirmation of the MD nature of this transition, we measure the excited-state lifetime of the lowest excited crystal field state for this site, $^4I_{13/2}(Y_1)$. The experimentally measured value, $\tau_0 = 21.04 \pm 0.02$ ms (Fig.~\ref{fig:establishMD}(c)) is considerably longer than \er lifetimes in many other materials (typically 5-10~ms \cite{stevenson2022}) and only slightly shorter than the theoretically predicted MD lifetime of $23.1$ ms, calculated from the cubic crystal-field model and the refractive index of MgO (see Supplementary Information). Therefore, we conclude that the overall $^4I_{13/2}(Y_1)$ decay is approximately 91\% MD, with the remainder being nonradiative or forced ED arising from a small distortion of the crystal. We note that other emitters in MgO have been observed to have significant nonradiative or forced ED phonon sideband transitions~\cite{karaveli2013}; their absence for \er is a consequence of the isolated nature of the 4$f$ electrons. The contribution of the next-order multipole, the electric quadrupole (E2), is estimated to be $\sim 1\times 10^{-7}$ smaller than the MD rate for the $^4I_{13/2}$ state of \er (see Supplementary Information)~\cite{dodson2012}.

To study the magnetic Purcell effect, we fabricate silicon nanophotonic resonators from a silicon-on-insulator wafer, and bond them onto Er-implanted MgO crystals using a stamping process described previously~\cite{dibos2018}. Each device consists of an array of cavities evanescently coupled to a single bus waveguide (Fig.~\ref{fig:fig1}(c)), which is coupled to an optical fiber using a grating coupler \cite{Chen2021a}. The in-plane distribution of the $B$ field is shown in Fig. 1(d); the peak field strength with a single photon in the cavity is $B_z = 8.4$ G. While the field decays exponentially into the substrate, it remains larger than 2.50~G at depths up to 100 nm.

In a first experiment to probe the magnetic Purcell enhancement, we use a sample implanted with $1\times 10^{12}$ Er/cm$^{2}$, distributed uniformly between the surface and a depth of 100 nm (sample B). After tuning the cavity resonance to the 1540.48 nm transition using gas deposition~\cite{dibos2018}, we probe the cavity-coupled ions using photoluminescence excitation spectroscopy (PLE) by sweeping a pulsed laser over the \er resonance while collecting time-delayed fluorescence photons with a superconducting nanowire single photon detector (SNSPD). PLE spectroscopy reveals a forest of single ion lines (Fig.~\ref{fig:single_ions}(a)). The single-ion nature is confirmed by measuring the second-order autocorrelation function ($g^{(2)}$) of the emitted photons (Fig. 3(c)). Individual ions have linewidths as narrow as 690 kHz (Fig. 3(b)), which is significantly smaller than previous measurements of shallow \er ions in Y$_2$SiO$_5$ \cite{dibos2018}, LiNbO$_3$ \cite{yang2023}, or silicon \cite{gritsch2023} (though we note that narrower linewidths for single \er ions have been observed for deeper ions in YSO~\cite{merkel2020}, and for shallow ions in a non-polar site in CaWO$_4$~\cite{ourari2023}). This is likely a consequence of the absence of a permanent electric dipole-moment for \er substituted at the cubic site, which renders the emitter insensitive to charge noise.

Focusing on a single ion, we determine the MD coupling strength to the cavity from the fluorescence lifetime. A representative time trace is shown in Fig. 3(d) alongside the bulk \ermgo{} fluorescence lifetime for comparison. The single ion decay rate is $\tau = 20.3 \pm 0.5$ $\mu$s, which is shorter than the bulk lifetime by $P_m = \tau_0/\tau = 1040 \pm 30$, thereby demonstrating strong magnetic Purcell enhancement. Using the relationship $P_m = 4 g_m^2/(\kappa \gamma)$, with the cavity decay rate $\kappa = 2\pi \times 3.14$ GHz ($Q = 6.2 \times 10^4$), we extract an atom-cavity coupling strength of $g_m = 2\pi \times 2.49$\,MHz. With $g_m = \mu B/\hbar$ and a theoretical dipole moment of $\mu = 0.62 \mu_B$ (where $\mu_B$ is the Bohr magneton), we determine the single-photon magnetic field of $B= 2.86$ G at the position of the ion. This is in good agreement with the simulated magnetic field strength for an optimally positioned ion at a depth of 65~nm (Fig.~1(d)), consistent with the expected ion distribution for sample B. 

Lastly, we use the Purcell-enhanced MD transition to realize a spin-photon interface. We use a third sample with an even lower \er implantation dose to allow clearer resolution of single ions (sample C, $2\times 10^{10}$ Er/cm$^{2}$). Based on the crystal field model, the ground state is expected to be a effective spin-3/2 quartet state ($\Gamma_8$ in Bethe notation), which is only allowed for \er at a cubic site \cite{Abragam1970}. However, an infinitesimal distortion breaking the cubic symmetry can lift this degeneracy, resulting in two Kramers' doublets, $Z_1$ and $Z_2$, and the presence of such a perturbation is suggested by the double-peak structure in Fig. 3(a), with a splitting of 1.5 GHz. A small magnetic field (50 G) further separates the Kramers' doublets into single states, that is, $Z_{1\pm}$ and $Z_{2\pm}$ (Fig.~4a).

With the laser and cavity tuned into resonance with the $Z_{1-} \rightarrow Y_{1-}$ transition, repeated application of laser pulses results in an exponential decay of the fluorescence (Fig.~(4b)). The initial fluorescence amplitude is proportional to the spin population in $Z_{1-}$, while the decay indicates optical pumping into other states. After initializing a population imbalance in this way, we were able to observe coherent Rabi oscillations between $Z_{1-}$ and $Z_{1+}$ while applying a microwave (MW) drive with frequency $f = 1.63$ GHz.

By using the initialization and readout sequence from Fig. 4(b), and calibrated microwave rotations from Fig. 4(c), we can measure the coherence properties of the $Z_1$ ground state. To measure the lifetime, we first initialize with optical pumping, and then measure the final state population in $Z_{1-}$ (using a direct fluorescence measurement) and $Z_{1+}$ (using a MW $\pi$ pulse followed by direct fluorescence measurement). We find a population relaxation time of 2.6 ms. In most other \er materials, the lowest Kramers' doublet has a relaxation time exceeding several seconds at sub-Kelvin temperatures~\cite{Raha2020,ourari2023}, because there is no direct phonon matrix element between the spin sublevels~\cite{Abragam1970}. We attribute the shorter observed lifetime to thermal transitions to the low-lying $Z_2$ level. If the separation of this level is much less than the Boltzmann energy $k_B T/h \approx 10$ GHz (here, $T=0.5 K$ is the environment temperature, $k_B$ is the Boltzmann constant, and $h$ is the Planck constant), then we expect the population of the lowest four levels $\{Z_{1-},Z_{1+},Z_{2-},Z_{2+} \}$ to change from approximately $\{0,0.33,0.33,0.33\}$ after optical pumping (assuming the $Y_1$ branching ratio to $Z_1$ and $Z_2$ is similar), to $\{0.25,0.25,0.25,0.25\}$ after thermalization. This is in agreement with the observed change in population of the $Z_{1+}$ state, providing support for the role of the $Z_2$ levels in the spin relaxation, although we note that we did not directly observe the $Z_2$ levels for the single ion. A similar relaxation timescale has been observed for \ermgo{} ensembles in bulk EPR measurements~\cite{Borg1970,Baker1976}.

Finally, we measure the spin coherence of the $\ket{Z_{1\pm}}$ states using Ramsey and Hahn echo sequences (Fig.~\ref{fig:spin}(e-f)), finding $T_2^* = 54 \pm 6$~ns and $T_2^{\textrm{Hahn}} = 560 \pm 40$~ns, respectively. The measured value of $T_2^*$ is in good agreement with the predicted range of values from the $^{25}$Mg nuclear spins bath ($I=5/2, g_n=-0.34$, relative abundance = 10\%), estimated using cluster correlation expansion (CCE)~\cite{Yang2008} simulations. However, the Hahn echo time, $T_2$, is shorter than predicted from the nuclear spin dynamics alone, indicating an additional source of noise with fast fluctuations. The most likely sources are native defects (that is, F centers \cite{hutchison1949}), paramagnetic impurities or surface spins.

In conclusion, we have demonstrated magnetic Purcell enhancement with a Purcell factor of $P_m = 1040 \pm 30$. This was enabled by engineering a new single photon emitter, Er$^{3+}$:MgO, and coupling it to a silicon nanophotonic cavity. This large Purcell enhancement establishes the equivalence of electric and magnetic density of states for nanophotonic engineering and opens the door to engineering strong light-matter interaction with a wider range of emitters.

There are several aspects of these results worth discussing. First, the observed magnetic Purcell factor $P_m = 1040 \pm 30$ is not only the first strong Purcell enhancement for an optical magnetic dipole transition, but it is also comparable to the largest Purcell factors ever reported for electric dipole transitions in any context, using nanopohtonic cavities~\cite{dibos2018,ourari2023} or plasmonic devices~\cite{russell2012,hoang2016}. This provides a striking demonstration of the equivalence of optical density of states engineering with electric and magnetic fields.

A particularly valuable application of magnetic Purcell enhancement is to develop new quantum emitters and spin-photon interfaces based on magnetic dipole transitions. Rare earth ions in centrosymmetric sites are one interesting class of MD emitters, which includes the example studied in this work. The main benefit of a centrosymmetric site is the absence of a permanent electric dipole moment and corresponding linear DC Stark shift in the presence of charge noise, which can suppress the spectral diffusion that prevents indistinguishable photon generation~\cite{ourari2023}, which is important for applications in quantum networks~\cite{pompli2021}. In fact, we attribute the narrow single-ion optical linewidth observed in this work to this effect. The observed nearly-quartet ground state is problematic for use as a qubit, because of fast spin relaxation. This effect could be overcome by the deliberate introduction of strain to further separate the $Z_1$ and $Z_2$ levels (\emph{e.g.}, using mechanical structures~\cite{meesala2018} or epitaxial growth~\cite{niu2006}), or by using host crystals that are centrosymmetric but not cubic. Finally, materials with strong optical MD transitions are of particular interest for the development of negative refractive index materials \cite{sikes2011,buckholtz2020}.

\emph{Acknowledgements} This work was primarily supported by the U.S. Department of Energy, Office of Science, National Quantum Information Science Research Centers, Co-design Center for Quantum Advantage (C2QA) under contract number DE-SC0012704. We also acknowledge support from the DOE Early Career award (for modeling of decoherence mechanisms and spin interactions), as well as AFOSR (FA9550-18-1-0334 and YIP FA9550-18-1-0081), the Eric and Wendy Schmidt Transformative Technology Fund, the Princeton Catalysis Initiative, and DARPA DRINQS (D18AC00015) for establishing the materials spectroscopy pipeline and developing integrated nanophotonic devices. CMP was supported by a National Defense Science and Engineering Graduate (NDSEG) Fellowship.

\bibliography{library.bib}

\clearpage

\begin{figure*}[ht]
    \includegraphics[width=6.5in]{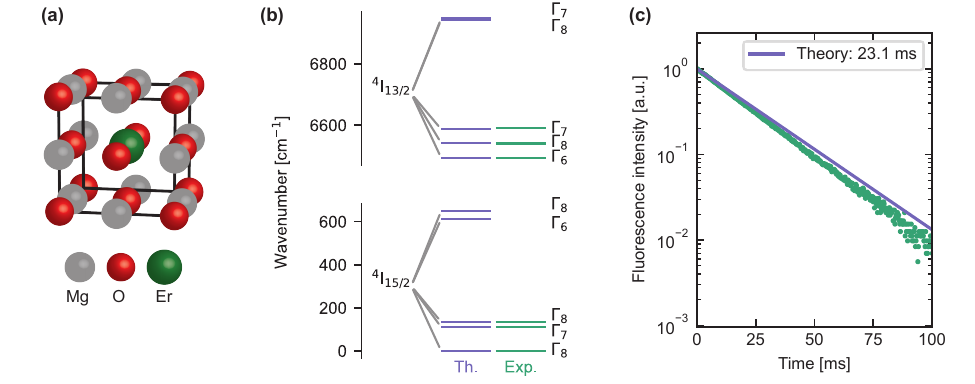}
    \caption{\label{fig:establishMD} \textbf{Establishing the magnetic nature of the 1.5 $\mu$m transition} (a) The cubic crystal structure of MgO, showing an Er$^{3+}$ ion substituted for Mg$^{2+}$. The indicated site is centrosymmetric; therefore, electric dipole transitions between $4f$ levels are strictly forbidden, resulting in a nearly pure magnetic dipole decay pathway. (b) Comparison of theoretical (purple) and experimental (green) energy levels of the $^4$I$_{15/2}$ and $^4$I$_{13/2}$ in the cubic site, measured in an \er ensemble. The close agreement corroborates the assignment of the cubic site. (c) Comparison of the predicted magnetic dipole decay rate (purple, 23.1 ms) and measured fluorescence lifetime (green, $21.04 \pm 0.02$~ms) of the $^4$I$_{15/2}(Z_1) \to ^4$I$_{13/2}(Y_1)$ transition. The close agreement suggests that the $^4$I$_{13/2}(Y_1)$ decay is more than 90\% magnetic dipole in nature.
    }

\end{figure*}

\begin{figure*}[ht]
    \includegraphics[width=6.5in]{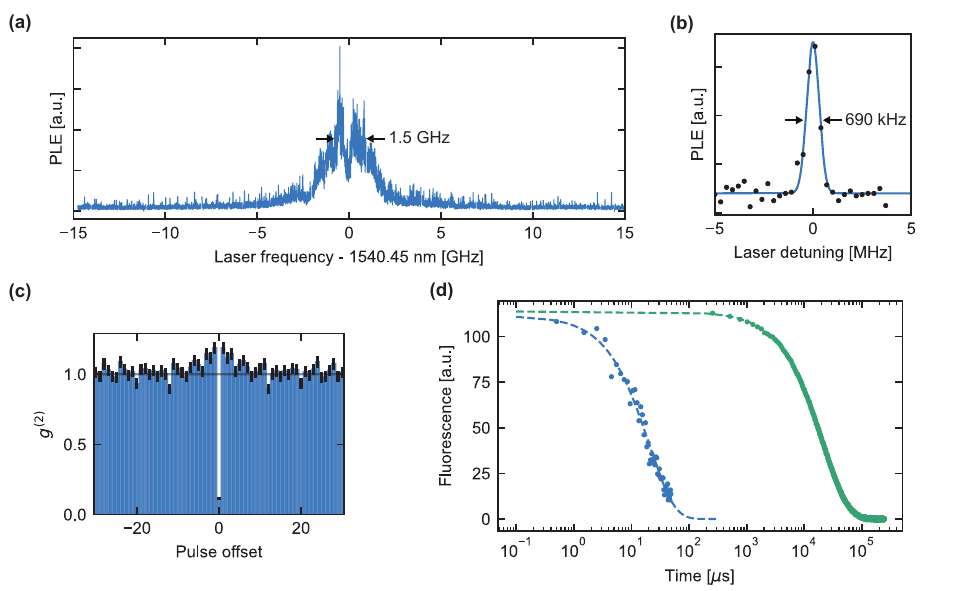}
    \caption{\textbf{Purcell enhancement of an optical magnetic dipole transition} (a) Photoluminescence excitation (PLE) spectrum of the ions coupled to one photonic crystal cavity. Resolved peaks corresponding to the transitions of individual ions are visible in the tails of the inhomogeneous distribution. (b) A representative isolated \er line, with a linewidth of 690 kHz. (c) Measurement of the second-order autocorrelation function of fluorescence photons from a single ion. At zero delay, $g^{(2)}(0) = 0.11 \pm 0.01$, confirming the single-emitter nature. (d) Fluorescence lifetime of a single cavity-coupled ion (blue). The data is fitted to an exponential decay with a time constant of $\tau = 20.3 \pm 0.5$~$\mu$s. For comparison, we also show the ensemble lifetime of a similarly prepared sample without a nanophotonic cavity (green, amplitude re-scaled), fit to a time constant of $\tau_0 = 21.04 \pm 0.02$~ms. From their ratio, we extract the magnetic Purcell enhancement factor $P_m = 1040 \pm 30$.}

    \label{fig:single_ions}
\end{figure*}

\begin{figure*}[ht]
    \includegraphics[width=6.5 in]{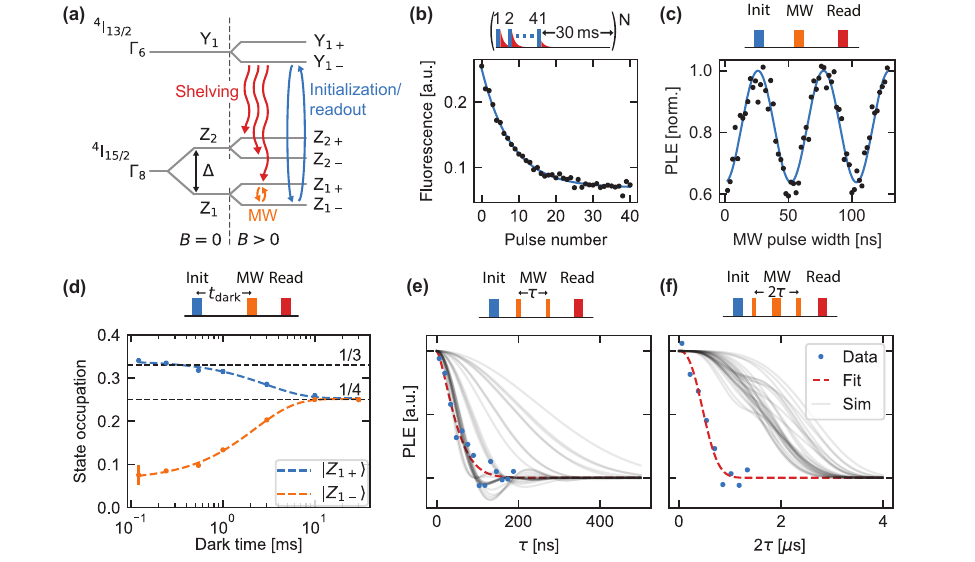}
    \caption{\textbf{Probing the spin dynamics of \ermgo{} using a magnetic dipole spin-photon interface} (a) Schematic diagram of relevant \ermgo{} energy levels and transitions. The four-fold degenerate ground state expected for cubic symmetry is split into two Kramers' doublets ($Z_1$, $Z_2$) by a lattice distortion, and then again into single levels by a magnetic field. The excited state is a doublet even in the absence of a lattice perturbation. Arrows indicate the optical excitation used to initialize and readout the spin (blue), other decay pathways leading to optical pumping (red), and the spin transition driven by microwaves (orange). (b) Repeated excitation of the $Z_{1-} \rightarrow Y_{1-}$ transition optically pumps the atom. The initial fluorescence level is also a measure of the initial spin population in $Z_{1-}$. (c) Rabi oscillations between $Z_{1-}$ and $Z_{1+}$, using a microwave pulse with frequency $f= 1.63$ GHz. (d) Spin relaxation following optical pumping out of $Z_{1-}$. After a variable wait time, the population in $Z_{1-}$ is measured using PLE; $Z_{1+}$ is also measured by inserting a microwave $\pi$ pulse before the readout. The population in each sublevel decays to the same equilibrium value with a time constant of $T_1 = 2.6 \pm 0.3$ ms. Based on the level diagram in panel a), the equilibrium population is taken to be 1/4, which defines the scale of the vertical axis. (e) Ramsey measurement of the spin coherence, yielding $T_2^*= 54 \pm 6$ ns. The gray lines show the prediction of a CCE-1 simulation of random configurations of $^{25}$Mg spins. (f) Hahn echo measurement of $T_2^{\text{Hahn}} = 560 \pm 40$ ns. The gray lines show CCE-2 simulations of the nuclear spin noise, suggesting that another decoherence mechanism dominates on longer timescales.}
    \label{fig:spin}
\end{figure*}

\clearpage

\section{Supplementary Information}
\section{\texorpdfstring{M\MakeLowercase{g}O}{MgO} sample preparation}

The MgO samples used in this study were procured from MTI Corporation. They have a double-sided epi polish, and are specified by the vendor to have a chemical purity $>99.95$ \%. Erbium was introduced into the samples using ion implantation (II-VI Inc.). A total of three different implantation and annealing treatments were used for different experiments. Sample A was implanted with an erbium density of $1\times 10^{14}$ Er/cm$^{2}$ targeting a uniform erbium distribution between the surface and a depth of 100 nm, using the fluences and energies shown in Tab.~\ref{tab:fluences}. After implantation, this sample was annealed in air at 600 $^\circ$C for 6 hours. Sample B was implanted with an erbium density of $1\times 10^{12}$ Er/cm$^{2}$ and targeted a uniform erbium density between the surface and a depth of 100 nm (Tab.~\ref{tab:fluences}), with no post-implantation anneal. Finally, sample C was implanted with an erbium density of $2\times 10^{10}$ Er/cm$^{2}$ using a 35 keV implantation energy corresponding to a target depth of 15 nm. After implantation, this sample was annealed in air at 400 $^\circ$C for 8 hours. The implantation energies were chosen to match the target depth based on simulations using the Stopping-Range of Ions in Matter package~\cite{ziegler2010}.
\begin{table}[ht!]
  \centering
  \begin{tabular}{rrrr}
    \hline
    \multicolumn{2}{c}{$1\times10^{12}$ Er/cm$^{2}$ total flux} & \multicolumn{2}{c}{$1\times10^{14}$ Er/cm$^{2}$ total flux} \\
    \hline
    Energy (keV) & Flux (Er/cm$^{2}$) & Energy (keV) & Flux (Er/cm$^{2}$) \\
    \hline 
    10 & $3.75 \times 10^{10}$ & 10 & $3.75 \times 10^{12}$ \\
    25 & $5 \times 10^{10}$ & 25 & $5 \times 10^{12}$ \\
    50 & $7.5 \times 10^{10}$ & 50 & $7.5 \times 10^{12}$ \\
    100 & $1 \times 10^{11}$ & 100 & $1 \times 10^{13}$ \\
    150 & $1.25 \times 10^{11}$ & 150 & $1.25 \times 10^{13}$ \\
    250 & $1.25 \times 10^{11}$ & 250 & $1.25 \times 10^{13}$ \\
    350 & $3.75 \times 10^{11}$ & 350 & $3.75 \times 10^{13}$ \\
    \hline
  \end{tabular}
  \caption{\label{tab:fluences} 
  Implantation schedule used to attain a uniform erbium distribution between the MgO surface and a target depth of 100~nm. Conditions for sample A are given in the $1\times10^{14}$ Er/cm$^{2}$ total flux column, whereas sample B details are in the $1\times10^{12}$ Er/cm$^{2}$ total flux column. }  
 \end{table}

\section{Experimental details}

Nanophotonic devices were fabricated from silicon on insulator wafers and transferred to the MgO substrate. Light was coupled to the devices via a grating coupler using an angle polished fiber and MW delivery was achieved using a scanning probe head. Details of the fabrication procedure and experimental apparatus can be found in Ref. \cite{Chen2021a}. 

The single ion fluorescence measurement (Fig.~3(a)) was performed at 4~K with the sample cooled with a Montana Inc Cryostation. All other single ion measurements were conducted at 500 mK using a BlueFors $^3$He cryostat.

The fluorescence measurement (Fig.~3(a)) was performed using 4.2~$\mu$s excitation pulses interleaved with 101 $\mu$s fluorescence collection windows, repeated $1\times10^4$ times for each frequency step, respectively. The narrow single ion measurement (Fig.~3(b)) utilized 4.2~$\mu$s excitation pulses and 41~$\mu$s fluorescence collection windows repeated $4\times10^5$ times. The second order auto-correlation experiment (Fig.~3(c)) used 4.2~$\mu$s excitation pulses and 500~$\mu$s fluorescence collection windows, repeated $7.36\times10^6$ times. The single ion lifetime measurement (Fig.~3(d)) utilized 4.2~$\mu$s excitation pulses interleaved with 101 $\mu$s fluorescence collection windows, repeated $2\times10^4$ times per shot. The lifetime was averaged over a total of 91 shots. 

Spin characterization experiments were all performed with sample C using two distinct magnetic field orientations. Initialization experiments used an external magnetic field magnitude of $50$~G and an orientation of $(\theta, \phi) = (60^{\text{o}}, 65^{\text{o}})$ selected to maximize cyclicity (see Fig.~1(c) of main text for coordinate system definition). Using this configuration, 41 pulses were used to initialize the spin state (Fig.~4(b) of main text). Increasing the number of pulses beyond this did not improve the initialization since an equilibrium between the optical pumping rate and phonon cross-relaxation rate (discussed in the main text) is reached. The same field configuration was utilized for spin $T_1$ characterization. The optical pulse sequence used for Fig.~4(b) and Fig.~4(d) consisted of a 4.2 $\mu$s optical excitation pulse interleaved with $61$~$\mu$s fluorescence collection windows, and was repeated a total of $1\times 10^4$ times. 

Rabi, Ramsey and Hahn echo measurements were performed using a magnetic field magnitude of $125$~G, with an orientation of $(\theta, \phi) = (90^{\text{o}}, 38^{\text{o}})$, selected to minimize background fluorescence from neighboring ions. This field corresponded to a ground state splitting of $f = 1.63$ GHz. The optical pulses used for acquiring Rabi oscillation data consisted of a 4.2~$\mu$s excitation pulses in tandem with 101~$\mu$s fluorescence windows, with each point repeated $4 \times 10^4$ times. The Ramsey data used optical excitation pulses of length 4.2 $\mu$s with fluorescence collection windows of 101~$\mu$s length, repeated $4\times10^4$ times for each data point. Finally, the Hahn data used optical excitation pulses of length $47$~ns interleaved with 101~$\mu$s length fluorescence collection windows and a total of $4.2 \times 10^5$ repetitions. The very short optical excitation pulses corresponded to calibrated optical $\pi$ pulses and were performed with a peak optical power four orders of magnitude larger than the uncalibrated 4.2~$\mu$s excitation pulses used in other experiments. 

For ensemble spectroscopy (Fig.~2(b-c) and Fig.~S1), we used a separate setup at 4~K, based on an Oxford Instruments Optistat cryostat. 
 
\section{Crystal-field modeling of the cubic  site}

A crystal field model was developed to verify that the observed transitions are consistent with a cubic point-group symmetry, and to determine a theoretical lifetime of the optical excited state.  
The complete Hamiltonian for the $4f$ electrons has the form
\begin{equation}
   H = H_{\mathrm{FI}} + H_{\mathrm{CF}} + H_{\mathrm{NMD}}, 
   \label{eqn:h_sum}
\end{equation}
where $H_{\mathrm{FI}}$ is the free-ion contribution, $H_{\mathrm{CF}}$ is the crystal-field Hamiltonian, and $H_{\mathrm{NMD}}$ is the nuclear magnetic-dipole interaction. The free-ion Hamiltonian follows the parameterization of Carnall \emph{et al.}~\cite{carnall1989}
\begin{equation}
   H_{\mathrm{FI}} = E_{\mathrm{AVG}} + \sum_{1,2,3} F^k f_k + \zeta_{4f} A_{\mathrm{SO}} + \alpha L(L+1) + \beta G(G_2) + \gamma G(R_7) + \sum_{i = 2,3,4,6,7,8} T^i t_i,
   \label{eqn:h_fi_defn}
\end{equation}
with $E_{\mathrm{AVG}}$ the central field Hamiltonian. Subsequent terms are arranged in pairs of parameters multiplied by the matrix elements of an operator. Specifically, $F^k$ are the Slater parameters and $f_k$ the components of the angular part of the electrostatic repulsion, $\zeta_{4f}$ is the spin-orbit coupling constant with $A_{\mathrm{SO}}$ the spin-orbit coupling operator. The parameters $\alpha$, $\beta$, and $\gamma$ account for two-body interactions (Trees parameters) and the parameters $T^i$ account for three-body interactions (Judd parameters). Furthermore, $G(G_2)$ and $G(R_7)$ are the eigenvalues of the Casimir operators of the groups $G_2$ and $R_7$ \cite{wybourne1965}.
\begin{figure*}[ht]
    \includegraphics[width=6.5 in]{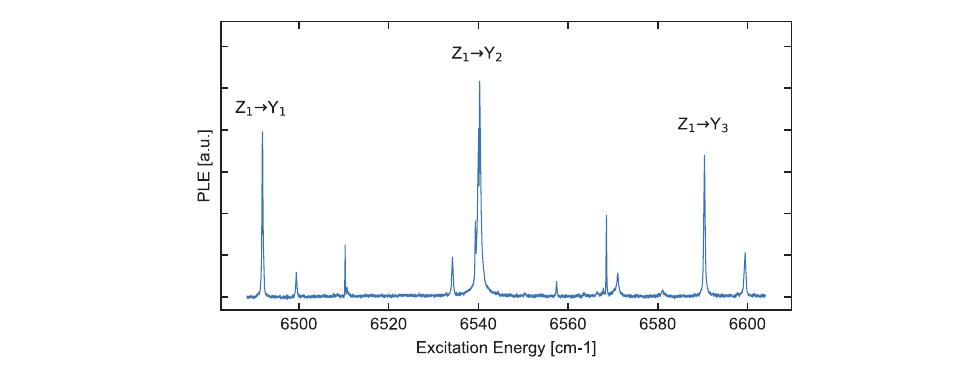}
    \caption{Excitation spectrum of \ermgo{} recorded using sample A at a temperature of 4~K. Transitions assigned to the cubic symmetry site have been labeled, whereas unlabeled sites correspond to additional spectroscopy sites. Additional spectroscopic data on other sites in \ermgo{} can be found in Ref.~\cite{stevenson2022}.}
    \label{fig:ex_spect}
\end{figure*}

For a substitutional site with a cubic point-group symmetry, the crystal-field Hamiltonian reads
\begin{equation}
   H_{\mathrm{CF}} = B^4_C \left[C^{4}_0 + \sqrt{\frac{5}{14}} \left(C^{(4)}_4 + C^{(4)}_{-4}\right)\right] + B^6_C \left[C^{6}_0 + \sqrt{\frac{7}{2}} \left(C^{(6)}_4 + C^{(6)}_{-4}\right)\right]. 
\end{equation}
 Here, the $B^4_C$ and $B^6_C$ are the crystal field parameters and $C^{(k)}_q$ are spherical tensor operators expressed using the normalization of Wybourne \cite{wybourne1965}. The last term in Eq.~\eqref{eqn:h_fi_defn}, $H_{\text{NMD}}$, accounts for the nuclear magnetic dipole; a detailed description of this contribution to $H$ can be found in Ref.~\cite{mcleod1997}.

The Hamiltonian parameters, shown in Tab.~\ref{tab:cf_param}, were optimized by fitting the eigenvalues of the Hamiltonian \eqref{eqn:h_sum} to the energy levels of \ermgo{} (Fig.~\ref{fig:ex_spect}) reported in Ref.~\cite{stevenson2022}.
\begin{table}[]
  \centering
  \begin{tabular}{rr}
    \hline
    Parameter & Value (cm$^{-1}$) \\
    \hline 
    $E_{\mathrm{AVG}}$    & 35879  \\ 
    $\zeta_{4f}$          & 2356   \\
    $B^4_C$               & 3403   \\
    $B^6_C$               & 290    \\
    $[F_2]$               & 97483  \\
    $[F_4]$               & 67904  \\
    $[F_6]$               & 54010  \\
    \hline
  \end{tabular}
  \caption{\label{tab:cf_param} 
  The Er$^{3+}$:MgO Hamiltonian parameters. Parameters in square brackets were held fixed during the optimization. The Slater parameters, along with all remaining parameters in Eq.\ \eqref{eqn:h_sum}, were fixed to the values reported for Er$^{3+}$:LaF$_3$ by Carnall et al. \cite{carnall1989}.}  
 \end{table}
The experimental and theoretical energies are summarized in Tab.~\ref{tab:energies}, yielding an r.m.s. difference of 1.6~cm$^{-1}$. This close agreement establishes that the studied \er site in \ermgo{} has cubic point-group symmetry. 

We note that due to available experimental data being limited to the $^4I_{15/2}$ and $^4I_{13/2}$ multiplets, it was only possible to fit the two crystal-field parameters $B^4_C$ and $B^6_C$, as well as the central field contribution $E_{\mathrm{AVG}}$ and the spin-orbit coupling constant $\zeta$. The remaining parameters were fixed to values of Er$^{3+}$:LaF$_3$ from Carnall \emph{et al.}~\cite{carnall1989}. As a consequence, the obtained fit is unlikely to accurately reproduce the inter-multiplet spacing of higher energy terms. However this does not affect the accuracy of the fitted cubic crystal-field parameters to leading order. Additionally, high-resolution spectroscopy of the $^4I_{15/2}(Z_1)$ to $^4I_{13/2}(Y_1)$ transition showed a lifting of the $\Gamma_8$ quartet into two Kramers' doublets with a splitting of 1.5~GHz. It was possible to reproduce this by adding a small axial perturbation to the cubic crystal field of $B^2_0 = 1.4$ cm$^{-1}$. To achieve close agreement with the observed lineshape, it was necessary to include an additional rank 2 contribution of $B^2_2 = 0.6$ cm$^{-1}$, along with the nuclear magnetic dipole hyperfine interaction with a coupling strength of 0.0054 cm$^{-1}$~\cite{horvath2019}. 
\begin{table}[]
    \centering 
    \begin{tabular}{lcS[table-format=4.1]}
    \hline
    State & Experiment (cm$^{-1}$) & {Cubic model (cm$^{-1}$)} \\
    \hline 
    $^4I_{15/2}(Z_{1,2})$* & 0, 0.08 & 0.0 \\
    $^4I_{15/2}(Z_3)$ & 110.4 & 110.4 \\
    $^4I_{15/2}(Z_{4,5})$* & 134.8 & 134.8 \\
    $^4I_{15/2}(Z_6)$ & - & 613.9 \\
    $^4I_{15/2}(Z_{7,8})$* & - & 651.9 \\
    & & \\
    $^4I_{13/2}(Y_1)$ & 6491.5 & 6491.8 \\
    $^4I_{13/2}(Y_{2,3})$* & 6539.8, 6540.1 & 6542.6 \\
    $^4I_{13/2}(Y_4)$ & 6590.2 & 6587.6 \\
    $^4I_{13/2}(Y_{5,6})$* & - & 6942.6 \\
    $^4I_{13/2}(Y_7)$ & - & 6945.7 \\
    \hline
    \end{tabular}
    \caption{\label{tab:energies}The Er$^{3+}$:MgO energy levels from Ref.~\cite{stevenson2022} and the corresponding theoretical energies calculated using the cubic crystal-field model. States that are theoretically predicted to correspond to $\Gamma_8$ quartets are marked with a star (*). The occurrence of three and two $\Gamma_8$ states for $J = 15/2$ and $J = 13/2$ multiplets, respectively, is consistent with the group theoretical prediction for an ion with an odd number of electrons in a cubic crystal-field \cite{dieke1968}. Two of the quartet states ($Z_{1,2}$ and $Y_{2,3}$) are experimentally observed to be split into very closely spaced doublets, consistent with a small distortion of the \er site.}
\end{table}

\section{Predicted magnetic dipole decay rate}
In order to evaluate the optical lifetime of the $^4I_{13/2}(Y_1)$ state for the cubic site, we use the Hamiltonian \eqref{eqn:h_sum} to determine the magnetic dipole moment. Following the treatment of Ref.~\cite{reid2006}, we note that
\begin{equation}
    {\boldsymbol \mu}^{(1)}_q = -\frac{e \hbar}{2 m_e c} (\mathbf{L} + 2 \mathbf{S})^{(1)}_q,
    \label{eqn:dipole_moment}
\end{equation}
where ${\boldsymbol \mu}^{(1)}_q$ is the rank 1 magnetic dipole operator with polarization $q = \{-1, 0, 1\}$. The transition strength for magnetic dipole transitions from initial state $I$ to final state $F$ can be found by summing over all the $i$ and $f$ components of these states:
\begin{equation}
    S^{\text{MD}}_{FI,q} = \sum_i \sum_f \vert \bra{Ff} {\boldsymbol \mu}^{(1)}_q \ket{Ii} \vert^2.
    \label{eqn:SMD}
\end{equation}
We note that the excited state lifetime is the inverse of the sum over emission rates from all polarizations. Therefore, one can define an effective emission rate by \cite{reid2006}
\begin{equation}
    A^\text{MD}_{FI} = \frac{1}{4 \pi \epsilon_0} \frac{4 \omega^3}{\hbar c^3} n^3 \frac{1}{\xi} \frac{1}{3} \sum_q S^{\text{MD}}_q. 
    \label{eqn:A_coeff}
\end{equation}
Here $\epsilon_0$ is the permittivity of free space, $\omega$ the angular frequency, $\hbar$ is Planck's constant, $c$ the speed of light, $n$ the refractive index of the host crystal, and $\xi$ the degeneracy of the initial state. 

The transition dipole moment ${\boldsymbol \mu}$ can be determined from the wavefunctions obtained from Eq.~\eqref{eqn:h_sum}. We find $\mu = 0.62 \mu_B$. For a refractive index of $n=1.715$ \cite{bond1965} we then find $A^{\text{MD}}=43.23$~s$^{-1}$, corresponding to a lifetime of $23.1$~ms. We note that the $^4I_{13/2}$ lifetime of \er has been previously calculated for the free-ion case in Ref~\cite{dodson2012}, yielding $19.6$~ms after scaling by $n^3$ to account for the MgO host. The slightly slower decay rate that we predict is related to the MD matrix element. We believe our predicted matrix element is more accurate because we include the effect of the crystal field ($H_{CF}$), and also fit the spin-orbit constant $\zeta$ to the spectroscopic data.

\section{Other contributions to the decay rate}

The theoretically calculated magnetic dipole relaxation rate for MgO is slightly slower than the experimentally observed decay rate, implying an additional relaxation path. It is therefore possible that the observed Purcell enhancement could be attributed to the other decay pathway, instead of the MD decay. However, given the comparable electric and magnetic mode volumes of the cavity, the maximum attainable Purcell factor for pure ED and pure MD emitters are also comparable. The observed Purcell factor agrees well with the predicted enhancement for a pure MD emitter. Therefore, it is not plausible to attribute the bulk of the Purcell enhancement to the < 10\% ED decay fraction.  

Finally we note that in addition to a small ED decay component, one may also consider the possibility of an electric quadrupole contribution. The electric quadrupole decay rate for the $^4I_{13/2}$ term of \er has been estimated using a free-ion Hamiltonian to be $\Gamma^{\text{EQ}}_{\text{FI}} = 1.16 \times 10^{-6}$~s$^{-1}$~\cite{dodson2012}. To first order, the free-ion case can be used to estimate the expected decay rate in a host material using $\Gamma^{\text{EQ}} = n^5 \Gamma^{\text{EQ}}_{\text{FI}}$~\cite{dodson2012}, which yields $\Gamma^{\text{EQ}} = 1.72 \times 10^{-5}$~s$^{-1}$. This decay rate is negligible compared to the MD decay rate and may therefore be disregarded.

\end{document}